\def \SAIT #1 #2 {{\em Mem.\ Soc.\ Astron.\ It.\/} {\bf #1}, #2}
\def \MESS #1 #2 {{\em The Messenger\/} {\bf #1}, #2}
\def \ASTRNACH #1 #2 {{\em Astron. Nach.\/} {\bf #1}, #2}
\def \AAP #1 #2 {{\em Astron. Astrophys.\/} {\bf #1}, #2}
\def \AAL #1 #2 {{\em Astron. Astrophys. Lett.\/} {\bf #1}, L#2}
\def \AAR #1 #2 {{\em Astron. Astrophys. Rev.\/} {\bf #1}, #2}
\def \AAS #1 #2 {{\em Astron. Astrophys. Suppl. Ser.\/} {\bf #1}, #2}
\def \AJ #1 #2 {{\em Astron. J.\/} {\bf #1}, #2}
\def \ANNREV #1 #2 {{\em Ann. Rev. Astron. Astrophys.\/} {\bf #1}, #2}
\def \APJ #1 #2 {{\em Astrophys. J.\/} {\bf #1}, #2}
\def \APJL #1 #2 {{\em Astrophys.. J. Lett.\/} {\bf #1}, L#2}
\def \APJS #1 #2 {{\em Astrophys. J. Suppl.\/} {\bf #1}, #2}
\def \APSS #1 #2 {{\em Astrophys. Space Sci.\/} {\bf #1}, #2}
\def \ASR #1 #2 {{\em Adv. Space Res.\/} {\bf #1}, #2}
\def \BAIC #1 #2 {{\em Bull. Astron. Inst. Czechosl.\/} {\bf #1}, #2}
\def \JSQRT #1 #2 {{\em J. Quant. Spectrosc. Radiat. Transfer\/} {\bf #1}, #2}
\def \MN #1 #2 {{\em Mon. Not. R. Astr. Soc.\/} {\bf #1}, #2}
\def \MEM #1 #2 {{\em Mem. R. Astr. Soc.\/} {\bf #1}, #2}
\def \PLR #1 #2 {{\em Phys. Lett. Rev.\/} {\bf #1}, #2}
\def \PASJ #1 #2 {{\em Publ. Astron. Soc. Japan\/} {\bf #1}, #2}
\def \PASP #1 #2 {{\em Publ. Astr. Soc. Pacific\/} {\bf #1}, #2}
\def \NAT #1 #2 {{\em Nature\/} {\bf #1}, #2}

\documentstyle{memsait}
\input epsf.sty

\def\lesssim{\mathrel{\hbox{\rlap{\hbox{\lower4pt\hbox{$\sim$}}}\hbox{$<$}}}}
\def\gtrsim{\mathrel{\hbox{\rlap{\hbox{\lower4pt\hbox{$\sim$}}}\hbox{$>$}}}}

\begin{opening}
\title{ ULTRA HIGH ENERGY COSMIC RAY and \\
UHE $\nu$-$\nu_r$-$Z$ Showering IN DARK HALOS}
\author{Daniele Fargion$^1$, P.G. De Sanctis Lucentini, M. Grossi, M. De Santis, Barbara Mele$^1$}
\institute{$^1$Physics Department and INFN,Rome,Italy\\
}
\date{} % DO NOT INSERT ANY DATE HERE !!!
\end{opening}

\begin{document}
%%%%%%%%%%%%%%%%%%%%%%%%%%%%%%%%%%%%%%%%%%%%%%%%%%%%%%%%%%%%%%%%%%%%%%%%%%%

\oddpagefooter{}{}{} % LEAVE AS IT IS !
\evenpagefooter{}{}{} % LEAVE AS IT IS !
\
\bigskip

%\vspace{1.4cm}
 % \centerline{Physics Department and INFN }
%  \centerline{\elevenrm Rome University 1,
%     Pl.A.Moro 2, 00185, Rome, Italy}
%\vspace{3cm}
\begin{abstract}
The Ultra High Energy Cosmic Ray (UHECR), by $\nu_r$-Z showering
in Hot Dark Halos (HDM), shows an energy spectra, an anisotropy
following the relic neutrino masses and  clustering in dark halo.
The lighter are the relic $\nu$ masses, the higher their
corresponding Z resonance energy peaks. A {\em twin }light
neutrino mass splitting may reflect a {\em twin } Z resonance and
 a complex UHECR spectra modulation ({\em a twin }) bump
at highest GZK energy cut-off.  Each possible $\nu$ mass
associates a characteristic dark halo size (galactic, local,
super cluster) and its anisotropy due to our peculiar position
within that dark matter distribution. The expected Z or WW,ZZ
showering into $p$ $\bar{p}$ and $n$ $\bar{n}$ should correspond
to peculiar clustering in observed UHECR at $10^{19}$,$2\cdot
10^{19}$, $4 \cdot 10^{19}$. A $\nu$ HDM halo around a Mpc will
allow to the UHECR $n$ $\bar{n}$ secondary component at $E_n>
10^{20}$ eV (due to Z decay) to arise playing a role comparable
with the charged $p$ $\bar{p}$ ones. Their un-deflected $n$
$\bar{n}$  flight is shorter leading to a prompt and hard UHECR
trace pointing toward the original UHECR source direction. The
direct $p$ $\bar{p}$ pairs are split and spread by random
magnetic fields into a more diluted and smeared and lower energy
UHECR signal around the original source direction. Additional
prompt TeVs signals by synchrotron radiation of electro-magnetic
Z showering must also occur solving the Infrared-TeV cut-off. The
observed hard doublet and triplets spectra, their time and space
clustering already favor the rising key role of UHECR $n$
$\bar{n}$ secondaries originated by $\nu$-Z tail shower.
\end{abstract}
\vspace{2.0cm}

%\twocolumn
\section{Introduction}
  Light Neutrino  may play a relevant role in
 Hot Dark Matter models within a hot-cold dark matter (HCDM) scenario.
 Their clustering in Galactic, Local dark halos offer the
possibility to overcome the Cosmic Black Body opacity ($\gtrsim 4
\cdot 10^{19}\,eV$) (GZK)  at highest energy cosmic ray
astrophysics. These rare events almost  in isotropic spread are
probably originated by blazars AGN, QSRs  in standard scenario,
and they should not come, if originally of hadronic nature, from
large distances (above tens Mpcs) because of the electro-magnetic
dragging friction of cosmic 2.75 K BBR and of the inter-galactic
radio backgrounds (GZK cut-off). Indeed as noted by Greisen,
Zatsepin and Kuzmin (K.Greisen 1966, Zat'sepin 1966), proton and
nucleon mean free path at E $> 5 \cdot 10^{19} \,EeV$ is less
than 30 $Mpc$ and asymptotically nearly ten $Mpc$.; also gamma
rays at those energies have even shorter interaction length ($10
\,Mpc$) due to severe opacity by electron pair production via
microwave and radio background interactions (R.J.Protheroe 1997).
Nevertheless these powerful sources (AGN, Quasars, GRBs)
suspected to be the unique source able to eject such UHECRs, are
rare  at nearby distances ($\lesssim 10 \div 20 \, Mpc$, as for
nearby $M87$ in Virgo cluster); moreover there are not nearby
$AGN$ in the observed UHECR arrival directions. Strong and
coherent galactic(R.J.Protheroe 1997) or extragalactic (Farrar et
all. 2000) magnetic fields, able to bend such UHECR (proton,
nuclei) directions are not really at hand. The needed coherent
lengths and strength are not easily compatible with known cosmic
magnetic fields. Finally in this scenario the $ZeV$ neutrons
born, by photo-pion proton conversions on BBR, may escape the
magnetic fields bending and should keep memory of the arrival
direction, leading to (unobserved)  clustering toward the primary
source. Secondaries EeV photons (by neutral pion decays) should
also abundantly point and cluster toward the same nearby $AGN$
sources (P.Bhattacharjee 2000, Elbert et all. 1995)  contrary to
$AGASA$ data. Another solution of the present GZK puzzle, the
Topological defects ($TD$), assumes as a source, relic heavy
particles of early Universe; they are imagined diffused as a Cold
Dark Matter component, in galactic or Local Group Halos.
Nevertheless the $TD$ fine tuned masses and ad-hoc decays are
unable to explain the growing evidences of doublets and triplets
clustering in $AGASA$~ $UHECR$ arrival data.  On the other side
there are growing evidences of self-correlation between UHECR
arrival directions with far Compact Blazars at cosmic distance
well above GZK cut-off (Tinyakov P.G.et Tkachev 2001). Therefore
the solution of UHECR puzzle based on primary Extreme High Energy
(EHE) neutrino beams (from AGN) at ZeV  $E_{\nu} > 10^{21}$ eV
and their undisturbed propagation from cosmic distances up to
nearby calorimeter (made by relic light $\nu$ in dark galactic or
local dark halo (Fargion et Salis 1997, Fargion, Mele et Salis
1999, Weiler 1999, Yoshida et all. 1998) is still, the most
favorite convincing solution for the GZK puzzle. New complex
scenarios for each neutrino mass spectra are then opening and
important signature of UHECR Z,WW
showering must manifest in observed anisotropy and space-time clustering.\\

\section{Relic $\nu_r$ neutrino masses and Hot Halo Clustering}

If relic neutrinos have a mass   larger than their thermal energy
(1.9 $K^0$) they may cluster in galactic or Local Group halos; at
eVs masses the clustering seem very plausible and it may play a
role in dark hot cosmology(Fargion 1983). Their scattering with
incoming extra-galactic EHE neutrinos determine high energy
particle cascades which could contribute or dominate the observed
UHECR flux at $GZK$ edges. Indeed the possibility that neutrino
share a little mass has been reinforced by Super-Kamiokande
evidence for atmospheric neutrino anomaly via $\nu_{\mu}
\leftrightarrow \nu_{\tau}$ oscillation. An additional evidence
of neutral lepton flavour mixing has been very recently reported
also by Solar neutrino experiment (SNO,Gallex,K2K). Consequently
there are at least two main extreme scenario for hot dark halos:
either $\nu_{\mu}\, , \, \nu_{\tau}$ are both extremely light
($m_{\nu_{\mu}} \sim m_{\nu_{\tau}} \sim \sqrt{(\Delta m)^2} \sim
0.05 \, eV$) and therefore hot dark neutrino halo is very wide,
smeared and spread out to local group clustering sizes
(increasing the radius but loosing in the neutrino density
clustering contrast), or $\nu_{\mu}, \nu_{\tau}$ may share
degenerated ($eV$ masses) split by a very tiny different values.
In the latter fine-tuned neutrino mass case ($m_{\nu}\sim 0.4
eV-1.2 eV$) (see Fig,2 and Fig.3) the Z peak $\nu \bar{\nu}_r$
interaction (Fargion et Salis 1997, Fargion, Mele et Salis 1999,
Weiler 1999, Yoshida et all. 1998) will be the favorite one; in
the second case (for heavier non constrained neutrino mass
($m_{\nu} \gtrsim 3 \, eV$)) only a $\nu \bar{\nu}_r \rightarrow
W^+W^-$(Fargion et Salis 1997, Fargion, Mele et Salis 1999), and
the additional $\nu \bar{\nu}_r \rightarrow ZZ$ interactions,
(see the cross-section in Fig.1)(Fargion et all. 2001) considered
here will be the only ones able to solve the GZK puzzle. Indeed
the relic neutrino mass within HDM models in galactic halo near
$m_{\nu}\sim 4 eV$, corresponds to a lower and $Z$ resonant
incoming energy

%\begin{subeqnarray}
\begin{equation}
%\sqrt{(\Delta m)^2}
 {{E_{\nu} =  {\left(
\frac{4eV} {\sqrt{{{m_{\nu}}^2+{p_{\nu}^2}}}} \right)} \cdot
10^{21} \,eV.} \nonumber}
\end{equation}
%\end{subeqnarray}

   This resonant incoming neutrino energy is unable to overcome GZK energies while it is
     showering mainly a  small energy fraction into nucleons ($p,\bar{p}, n, \bar{n}$),
   (see $Tab.1$ below), at energies $E_{p}$ quite below. (see $Tab.2$
   below).

\begin{equation}
%\sqrt{(\Delta m)^2}
 {{E_{p} =  2.2 {\left(
\frac{4eV} {\sqrt{{{m_{\nu}}^2+{p_{\nu}^2}}}} \right)} \cdot
10^{19} \,eV.} \nonumber}
\end{equation}

   Therefore too heavy ($> 1.5 eV$) neutrino mass are not fit to
   solve GZK by Z-resonance; on the contrary WW,ZZ showering as well as t-channel showering
   may naturally keep open the solution.
   In particular the overlapping of both the Z and the WW, ZZ
   channels described in fig.1, for $m_{\nu} \simeq 2.3 eV$ while
   solving the UHECR above GZK they must pile up (by Z-resonance
   peak activity) events at $ 5 \cdot 10^{19} eV$, leading to a bump in
   AGASA data. There is indeed a first marginal evidence of such a UHECR bump
    in AGASA and Yakutsk data that may stand for this interpretation.
     More detailed data are  needed to verify such very exciting  possibility.
      Similar result regarding the fine tuned relic mass
    at $0.4 eV$ and $2.3 eV$, (however ignoring the WW ZZ and
    t-channels and invoking very hard UHE neutrino spectra) have been
    independently reported recently (Fodor et all. 1999) .
   Most of us  consider cosmological light relic neutrinos in Standard Model
   at non relativistic regime   neglecting any relic neutrino momentum ${p_{\nu}} $ term.
     However, at lightest mass values  the momentum may be comparable to the relic mass;
     moreover the spectra may reflect additional relic neutrino-energy injection
     which are feeding   standard cosmic relic neutrino
        at energies much above the same neutrino mass.
    Indeed there may  exist, within or beyond Standard Cosmology,
      a relic neutrino component due to stellar,
   Super Nova, GRBs, AGN past activities, presently  red-shifted
    into a  KeV-eV spectra, piling into a relic neutrino grey-body  spectra.
     Therefore  it is worth-full to keep the most general
      mass and momentum term in the target relic neutrino spectra.
      In this windy ultra-relativistic neutrino  cosmology, eventually
      leading to a neutrino radiation dominated Universe, the
      halo size to be considered is nearly coincident with the GZK one defined by
      the energy loss lenght for UHECR nucleons ($\sim 20 Mpcs$).
       Therefore the isotropic UHECR behaviour
      is guaranteed but a puzzle related to uniform source
      distribution  seem to persist. Nevertheless the UHE neutrino-
      relic neutrino scattering \textit{do not} follow a flat
      spectra as shown in figure 2, (as well as any hypothetical $\nu$ grey
      body spectra). This leave  open the opportunity to have a
      relic relativistic neutrino component at eVs energies as
      well as the observed  non uniform UHECR spectra. This case is similar to the
       case of a very light neutrino mass much below $0.1$ eV.
As we noticed above, relic neutrino mass above a few eVs in  HDM
halo \textit{are not} consistent with naive Z peak; higher
energies interactions ruled by WW,(D.Fargion, B.Mele et A.Salis
1999; K.Enqvist et all. 1989) ZZ cross-sections (Fargion 2001) may
nevertheless solve the GZK cut-off. In this regime there will be
also possible to produce by virtual W exchange, t-channel, $UHE$
lepton pairs, by $\nu_i \bar{\nu}_j\rightarrow l_i\bar{l}_j$,
leading to additional electro-magnetic showers injection.\\
As we shall see these important and underestimated signal will
produce UHE electrons whose final trace are TeVs synchrotron
photons able to break the IR-TeV cosmic cut-off.
 The hadronic tail of the Z or $W^+ W^-$ cascade maybe
 the source of final  nucleons $p,\bar{p}, n, \bar{n}$ able to explain UHECR events observed by
Fly's Eye and AGASA (Y.Uchihori et al. 2000) and other detectors.
The same $\nu \bar{\nu}_r$ interactions are source of Z and W that
decay in rich shower ramification. The average energy deposition
for both gauge bosons among the secondary particles is summarized
in Table 1A below.

%%%%%%%%%%%%%%%   Figure 1 Ex 01  %%%%%%%%%%%%%%%%%   FFFFFFFFFFFFFFFFFFFFFFFFFFFFFFFFFFFFFF
\begin{figure}
\epsfysize=0.30\textwidth % fix the y-dimension and scales x-dim. to y-dim.
\hspace{0.05\textwidth}\epsfbox{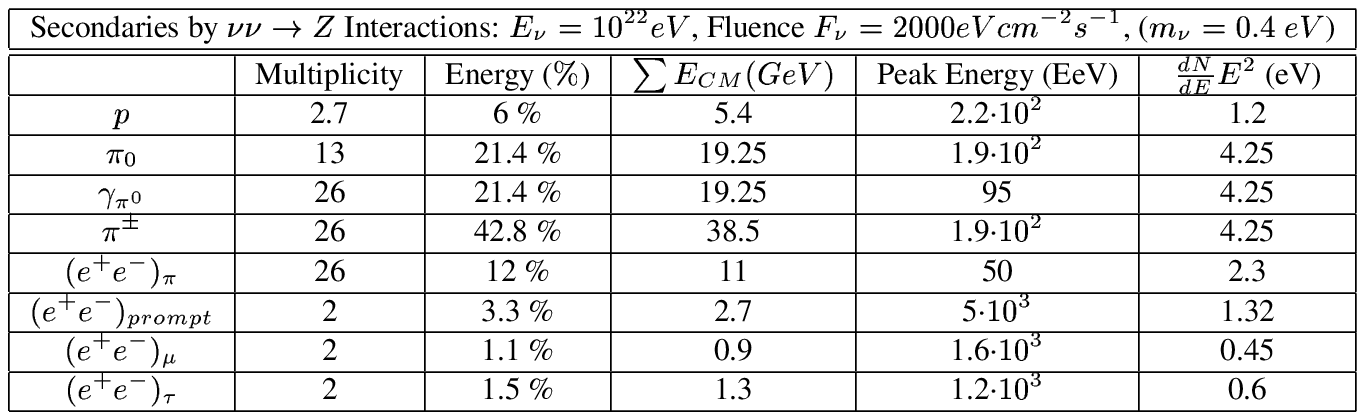} %for centering: act on hspace argument
\caption[h]{Table 1A: The  detailed energy percentage
distribution  into neutrino, protons, neutral and charged pions
and consequent gamma, electron pair particles both from hadronic
and leptonic Z, $WW,ZZ$ channels. We  calculated the
elecro-magnetic contribution due to the t-channel $\nu_i \nu_j$
interactions. We used LEP data for Z decay and considered W decay
roughly in the same way as Z one. We assumed that an average
number of 37 particles is produced during a Z (W) hadronic decay.
The number of prompt pions both charged (18) and neutral (9), in
the hadronic decay is increased by 8 and 4 respectively due to
the decay of $K^0$, $K^{\pm}$, $\rho$, $\omega$, and $\eta$
particles. (*)We assumed that the most energetic neutrinos
produced in the hadronic decay mainly come from charged pion
decay. So their number is roughly three times the number of
$\pi$'s. UHE photons are mainly relics of neutral pions. Most of
the $\gamma$ radiation will be degraded around PeV energies by
$\gamma \gamma$ pair production with cosmic 2.75 K BBR, or with
cosmic radio background. The electron pairs instead, are mainly
relics of charged pions and will rapidly lose energies into
synchrotron radiation. The contribution of leptonic Z (W) decay
is also considered and calculated in the table above and below.}
\end{figure}
%%%%%%%%%%%%%%%   Figure 1 END   %%%%%%%%%%%%%%%%%   FFFFFFFFFFFFFFFFFFFFFFFFFFFFFFFFFFFFFF

%%%%%%%%%%%%%%%   FFFFFFFFFFFFFFFFFFFFFFFFFFFFFFFFFFFFFF
%\begin{table}[h]
%\begin{center}
%\begin{tabular}{|c|c|c|c|}
%\hline
%     & Z & $W^+ W^- $ & t-channel \\ \hline
%  $\nu$ & 58 \% & 55 \% & 47 \% \\ \hline
%    $\gamma$
%& 21 \% & 21 \% & 4 \% \\ \hline
%    $e^+ e^-$ & 16 \% & 19 \% & 49 \% \\ \hline
%  $p$ & 5 \% & 5 \% & - \\ \hline
%\end{tabular}
%\end{center}
%\end{table}

\section{UHECR  channels in Z showers}

 Although protons (or anti-protons)
  are the most popular and favorite candidate in
order to explain the highest energy air shower observed, one
doesn't have to neglect the signature of final neutron and
anti-neutrons as well as electrons and photons. Indeed the UHECR
neutrons are produced in Z-WW showering at nearly same rate as
the charged nucleons. Above GZK cut-off energies UHE $n$,$
\bar{n}$, share a life lenght comparable with the Hot Galactic
Dark Neutrino Halo. Therefore they may be an important component
in UHECRs. Moreover prompt UHE electron (positron) interactions
with the galactic or extra-galactic magnetic field or soft
radiative backgrounds may lead to gamma cascades and from PeVs to
TeVs energies.\\
Gamma photons at energies $E_{\gamma} \simeq 10^{20}$ - $10^{19}
\,eV$ may freely propagate through galactic or local halo scales
(hundreds of kpc to few Mpc) and could also contribute to the
extreme edges of cosmic ray spectrum  and clustering
(Yoshida et al. 1998, Fargion et al. 2001). \\
The ratio of the final energy flux of nucleons near the Z peak
resonance, $\Phi_p$ over the corresponding electro-magnetic
energy flux $\Phi_{em}$ ratio is, as in tab.1 $e^+ e^-,\gamma$
entrance, nearly $\sim \frac{1}{8}$.  Moreover if one considers
at higher $E_{\nu}$ energies, the opening of WW, ZZ channels and
the six pairs $\nu_e \bar{\nu_{\mu}}$, \, $\nu_{\mu}
\bar{\nu_{\tau}}$, \, $\nu_e \bar{\nu_{\tau}}$ (and their
anti-particle pairs) t-channel interactions leading to  highest
energy leptons, with no nucleonic relics (as $p, \bar{p}$), this
additional injection favors the electro-magnetic flux $\Phi_{em}$
over the corresponding nuclear one $\Phi_p$ by a factor $\sim
1.6$ leading to $\frac{\Phi_p}{\Phi_{em}} \sim \frac{1}{13}$.
This ratio is valid at $WW,ZZ$ masses because the overall cross
section variability is energy dependent. At center of mass
energies above these values, the $\frac{\Phi_p}{\Phi_{em}}$
decreases more because the dominant role of t-channel (Fig1). We
focus here  on Z, and WW,ZZ channels showering in hadrons for GZK
events.
The important role of UHE electron showering into TeV radiation is discussed below.\\

%\section{UHE $\nu$ - $\nu_{relic}$ Cross Sections }

  There is an upper bound density clustering for
very light Dirac fermions due to the maximal Fermi degenerancy
whose adimensional density contrast is $\delta\rho \propto
m_{\nu}^3$,  while one finds (Fargion 1983)  that  the neutrino
free-streaming halo grows only as $\propto m_{\nu}^{-1}$.
Therefore the overall interaction probability grows $ \propto
m_{\nu}^{2} $, favoring heavier non relativistic (eVs) neutrino
masses. In this frame above few eV neutrino masses only WW-ZZ
channel are operative. Nevertheless the same lightest relic
neutrinos may share higher Local Group velocities (thousands
$\frac{Km}{s}$) or even nearly relativistic speeds and it may
therefore compensate the common density bound:

\begin{equation}
 n_{\nu_{i}}=1.9\cdot10^{3}%\left(\frac{n_{\nu_{cosmic}}}{54cm^{-3}}\right)
\left( \frac{m_{i}}{0.1eV}\right)  ^{3}\left(
\frac{v_{\nu_{i}}}{2\cdot10^{3}\frac{Km}{s} }\right)  ^{3}
\end{equation}

%\begin{equation} { {n_{\nu_{\i}}}= {10^3 {\left(
%\frac{n_{\nu_i}}{54\,{\rm cm}^{-3}}\right)} \;
%\left(\frac{m_{\nu_i}}{{\rm 0.1eV}}\right)^3 \left( \frac{\
%v_{\nu_i}}{2000{\rm km/s}}\right)^3\,\right}} \end{equation}

%\\

 From the cross section side there are three main interaction processes that
 have to be considered  leading to nucleons in the
of EHE and relic neutrinos scattering.

 {\bf channel 1.} $\;$ The
$\nu \nu_r\rightarrow Z \rightarrow \, $
 annihilation at the Z resonance.

{\bf channel 2.} $\nu_{\mu} \bar{\nu_{\mu}} \rightarrow W^+ W^-$
or $\nu_{\mu} \bar{\nu_{\mu}} \rightarrow Z Z$ leading to
hadrons, electrons, photons, through W and Z decay.

 {\bf channel 3.} The $\nu_e$ - $\bar{\nu_{\mu}}$, $\nu_e$ -
$\bar{\nu_{\tau}}$, $\nu_{\mu}$ - $\bar{\nu_{\tau}}$ and
antiparticle conjugate interactions of different flavor neutrinos
mediated in the $t$-channel by the W exchange (i.e. $\nu_{\mu}
\bar{\nu_{\tau_r}} \rightarrow \mu^- \tau^+ $). These reactions
are sources of prompt and secondary UHE electrons as well as
photons resulting by hadronic $\tau$ decay.

Their cross-section values are plotted in Fig.1.
 The asymptotic behaviour of
these cross section is proportional to
$\sim(\frac{M_W^2}{s})\ln{(\frac{s}{M_W^2})}$ for $s\gg M_Z^2$.\\
The nucleon arising from WW and ZZ hadronic decay could provide a
reasonable solution to the UHECR events above GZK. We'll assume
that the fraction of pions and nucleons related to the total
number of particles from the W boson decay is the almost the same
of Z boson. So W hadronic decay ($P \sim 0.68$) leads on average
to about 37 particles, where $<n_{\pi^0}> \sim 9.19$,
$<n_{\pi^{\pm}} > \sim 17$, and $<n_{p,\bar{p}, n, \bar{n}}> \sim
2.7$. In addition we have to expect by the subsequent decays of
$\pi$'s (charged and neutral), kaons and resonances ($\rho$,
$\omega$, $\eta$) produced, a flux of secondary UHE photons and
electrons. As we already pointed out, the particles resulting
from the decay are mostly prompt pions. The others are particles
whose final decay likely leads to charged and neutral pions as
well. As a consequence the electrons and photons come from prompt
pion decay.  On average it results (Fargion, Grossi et Lucentini
2001; Fargion, Grossi, Lucentini et Troia 2001) that the energy in
the bosons decay is not uniformly distributed among the particles.
Each charged pion will give an electron (or positron) and three
neutrinos, that will have less than one per cent of the initial W
boson energy, while each $\pi^0$ decays in two photons, each with
1 per cent of the initial W energy. In the Table~1A above we show
all the channels leading from single Z,W and Z pairs as well as
t-channel in nuclear and electro-magnetic components.
%Their energies and corresponding fluence are summirized in Table 2.\\
%At the same time a pure W leptonic decay $W \rightarrow l \nu$
%could occur for each flavor with a probability $P \sim 0.11$
%(see Table 4).\\

%%%%%%%%%%%%%%%   Figure 1 Ex 01  %%%%%%%%%%%%%%%%%   FFFFFFFFFFFFFFFFFFFFFFFFFFFFFFFFFFFFFF
\begin{figure}
\epsfysize=7cm
\hspace{2.0cm}\epsfbox{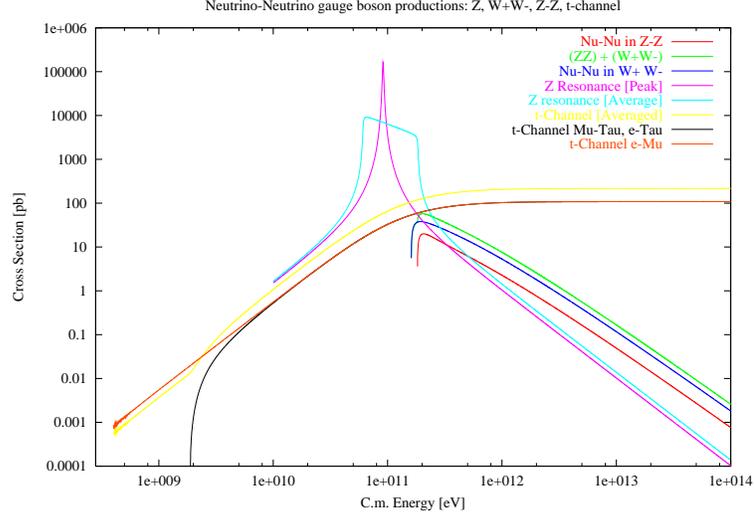} %for centering: act on hspace argument
\caption[h]{The  $\nu \bar{\nu} \rightarrow  Z,W^+
W^-,ZZ,T$-channel,  cross sections as a function of the center of
mass energy in $\nu \nu$.
   These cross-sections are estimated also in average (Z) as well for each possible
   t-channel lepton pairs. The averaged t-channel averaged the multiplicity of flavours
    pairs ${\nu}_{i}$, $ \bar{\nu}_{j}$ respect to neutrino
    pair annihilations into Z neutral boson. The Z-WW-ZZ Showering
    has to be relativistically boosted to show their behaviour at laboratory system.}
\end{figure}

%%%%%%%%%%%%%%%   Figure 1 END   %%
This interactions, as noted in Table~1A are leading to
electro-magnetic showers and are not offering any nuclear
secondary.

\section{The Boosted Z-UHECR  spectra }

Let us examine the destiny of UHE primary particles (nucleons,
electrons and photons) ($E_e \lesssim 10^{21}\,eV$) produced
after hadronic or leptonic W decay. As we already noticed in the
introduction, we'll assume that the nucleons, electrons and
photons spectra (coming from W or Z decay) after $\nu \nu$
scattering in the halo, follow a power law that in the center of
mass system is $\frac{dN^*}{dE^* dt^*} \simeq E^{* - \alpha}$
where $\alpha \sim 1.5$. This assumption is based on detailed
Monte Carlo simulation of a heavy fourth generation neutrino
annihilations(Yu. A.Golubkov 1998; D. Fargion,Yu.A.Golubkov,
M.Yu.Khlopov 1999; D. Fargion, R. Konoplich et all. 2000) and
with the model of quark - hadron fragmentation spectrum suggested
by Hill (C.T.Hill 1983).

In order to determine the shape of the particle spectrum in the
laboratory frame, we have to introduce the Lorentz relativistic
transformations from the center of mass system to the laboratory
system.
 The number of particles is clearly a relativistic invariant $dN_{lab} = dN^*$,
while the relation between the two time intervals is $dt_{lab} =
\gamma dt^*$, the energy changes like $ \epsilon_{lab} = \gamma
\epsilon^* (1 + \beta \cos \theta^*) = \epsilon^* \gamma^{-1}(1 -
\beta \cos \theta)^{-1}$, and finally the solid angle in the
laboratory frame of reference becomes $d\Omega_{lab} =\gamma^{2}
d\Omega^*  (1 - \beta \cos \theta )^2$. Substituting these
relations one obtains

%\begin{subeqnarray}
%\left(\frac{dN}{d\epsilon dt d\Omega} \right)_{lab} =
%\frac{dN_{*}}{d\epsilon_{*} dt_{*} d\Omega_{*}} \gamma^{-2}
% (1 - \beta \cos \theta)^{-1} =
% \frac{\epsilon^{-\alpha}_{*} \; \gamma^{-2}} {4 \pi}} \cdot (1 - \beta \cos
% \theta)^{-1} \nonumber \\
%\left( \frac{dN}{d\epsilon dt d\Omega} \right)_{lab} =
% \frac{\epsilon^{-\alpha} \; \gamma^{-\alpha-2}} {4 \pi}} (1 - \beta \cos \theta)^{-\alpha-1}\setcounter{eqsubcnt}{0}
%\end{subeqnarray}

\begin{displaymath}  %\begin{multline}
\left(\frac{dN}{d\epsilon dt d\Omega} \right)_{lab} =
\frac{dN_{*}}{d\epsilon_{*} dt_{*} d\Omega_{*}} \gamma^{-2}
 (1 - \beta \cos \theta)^{-1} =
 \end{displaymath}

\begin{equation}
= \frac{\epsilon^{-\alpha}_{*} \; \gamma^{-2}} {4 \pi} \cdot (1 -
\beta \cos \theta)^{-1} = \frac{\epsilon^{-\alpha} \;
\gamma^{-\alpha-2}} {4 \pi} (1 - \beta \cos \theta)^{-\alpha-1}
\end{equation}  %\end{multline}

and integrating on $\theta$ (omitting the lab notation) one loses
the spectrum dependence on the angle.

%\begin{equation}
%\left( \frac{dN}{d\epsilon dt d\Omega} \right)_{lab} \propto
%\epsilon^{-\alpha} \gamma_{Z (W)}^{ \alpha} \sim \epsilon^{-
%\frac{\alpha}{2}} \sim \epsilon^{- 0.75}.
%\end{equation}

The consequent fluence derived by the solid angle integral is:
% Long version
%\begin{equation}
% \frac{dN}{d\epsilon dt} \epsilon^{2}=
% \frac{\epsilon^{-\alpha+2} \; \gamma^{\alpha-2}} {2 \beta \alpha}}
% [(1 + \beta)^{\alpha} -
%  \frac{1} {[(1 + \beta)\gamma^2]^{\alpha}}}] \simeq
% \frac{\epsilon^{-\alpha+2} \; \gamma^{\alpha-2}} {\alpha}}
%\end{equation}

\begin{equation}  %\begin{multline}
%\begin{split}
\frac{dN}{d\epsilon dt} \epsilon^{2}= \\
 \frac{\epsilon^{-\alpha+2} \; \gamma^{\alpha-2}} {2 \beta \alpha}
 [(1 + \beta)^{\alpha} - (1 - \beta)^{\alpha}]
 \simeq \frac{2^{\alpha-1}\epsilon^{-\alpha+2} \; \gamma^{\alpha-2}} {\alpha}
%\end{split}
\end{equation}  %\end{multline}

There are two extreme case to be considered: the case where the
interaction occurs at Z peak resonance and therefore the center of
mass Lorents factor $\gamma$ is frozen at a given value (eq.1)
and the case (WW,ZZ pair channel) where all energies are
allowable  and $\gamma$ is proportional to $\epsilon^{1/2}$.
%; the latter case will be discussed in detail elsewhere.
 Here we focus only on Z peak resonance. The consequent fluence spectra
 $\frac{dN}{d\epsilon dt}\epsilon^{2}$, as above, is proportional to $\epsilon^{-\alpha +2}$. Because $\alpha$ is
nearly $1.5$ all the consequent secondary particles will also show
a spectra proportional to $\epsilon^{1/2}$ following a normalized
energies shown in Tab.2, as shown in Fig.(2-6). In the latter
case (WW,ZZ pair channel), the relativistic boost reflects on the
spectrum of the secondary particles, and the spectra power law
becomes $\propto \epsilon^{\alpha/2 +1}=\epsilon^{0.25}$. These
channels will be studied in details elsewhere. In Fig.~1 we show
the spectrum of protons, photons and electrons coming from Z
hadronic and leptonic decay assuming a nominal primary CR energy
flux $\sim 20~eV s^{-1} sr^{-1} cm^{-2}$, due to the total $\nu
\bar{\nu}$ scattering at GZK energies as shown in figures 2-6.
Let us remind that we assume an interaction probability of $\sim
1 \%$ and a corresponding UHE incoming neutrino energy $\sim
2000~eV s^{-1} sr^{-1} cm^{-2}$ near but below present $UHE$
neutrino flux bound from AMANDA and Baikal as well as Goldstone
data.

%\begin{center}
 \vspace{0.3cm}
\begin{table}[h]
\begin{center}
\begin{tabular}{ccc}
\hline
 \cline{1-3}
%\vspace{-0.6cm} &  & \\ \cline{1-3}
 % after \\ : \hline or \cline{col1-col2} \cline{col3-col4}...
%\multicolumn{3}{c}{{SECONDARIES ENERGY DISTRIBUTIONS}} \\
%\multicolumn{3}{c}{IN Z DECAY ($m_{\nu}=0.4 \; eV$)} \\
\multicolumn{3}{c}{\vspace{-0.2cm}Secondaries Energy Distributions}
\\
\multicolumn{3}{c}{
In Z Decay ($m_{\nu}=0.4 \; eV$)} \\
\hline \cline{1-3} \vspace{-0.15cm}\\
%%% \cline{1-3} %%%% \vspace{-0.6cm} &  & \\ \cline{1-3}
$~~~~Channel~~~~$ & $E (eV)$ & $\frac{dN}{dE}E^2$ (eV) \\
\vspace{-0.1 cm}\\ \cline{1-3}
  $p$ &  $2.2 \cdot 10^{20}$ & 1.2 \\ \cline{1-3}
  $\gamma$ & $9.5 \cdot 10^{19}$ & 4.25 \\ \cline{1-3}
  $e_{\pi}$ & $5 \cdot 10^{19}$ & 2.3 \\ \cline{1-3}
   $e_{prompt}$ & $5 \cdot 10^{21}$ & 1.32 \\ \cline{1-3}
    $e_{\mu}$ & $1.66 \cdot 10^{21}$ & 0.45 \\ \cline{1-3}
     $e_{\tau}$ & $1.2 \cdot 10^{21}$ & 0.6 \\ \cline{1-3}
\vspace{-0.8cm} &  & \\ \cline{1-3}
\end{tabular}
\end{center}

\caption{B. Summary of Energy peak and Energy Fluence for
different decay channels as described in Table 1A to build up the
Fig.1; $m_{\nu} = 0.4 eV$}
\end{table}

\section{The UHECRs from Relic $\nu$ Masses}

The role of each relic neutrino mass is summirized from the
convolutions of the UHE neutrino spectra with the relic neutrino
mass, its density as well as the cross-sections described above.
The case of Z-resonance event with a single neutrino mass has a
narrow fine tuned energy mass windows (0.4 eV-1.2 eV) described
respectively in Figures 3-4.

%%%%%%%%%%%%%%%   Figure 3       %%%%%%%%%%%%%%%%%   FFFFFFFFFFFFFFFFFFFFFFFFFFFFFFFFFFFFFF

\begin{figure}
\epsfysize=7cm \hspace{2 cm}\epsfbox{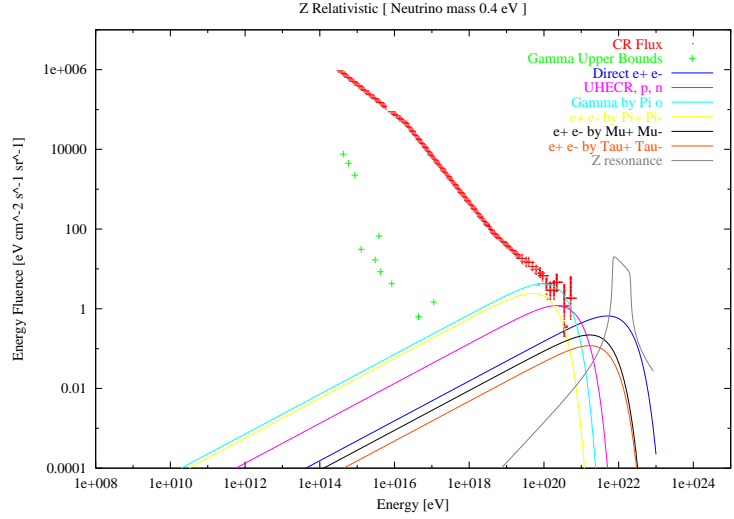} \caption[h]{Energy
Fluence derived by $\nu \bar{\nu} \rightarrow Z$ and its
showering into
  different channels: direct electron pairs UHECR nucleons $n$ $p$ and anti-nucleons, $\gamma$ by $\pi^0$ decay,
  electron pair by $\pi^+ \pi^-$ decay, electron pairs by direct muon and tau decays as labeled in figure.
  The relic neutrino mass has been assumed to be fine tuned to explain GZK UHECR tail:
  $m_{\nu}=0.4 eV$. The Z resonance ghost (the shadows of Z Showering resonance (Fargion 2001) curve),
  derived from Z cross-section
  in Fig.1, shows the averaged $Z$ resonant cross-section peaked
  at $E_{\nu}=10^{22} eV$. Each channel shower has been normalized following table 1B.}
\end{figure}

%%%%%%%%%%%%%%%   Figure 3 END   %%%%%%%%%%%%%%%%%   FFFFFFFFFFFFFFFFFFFFFFFFFFFFFFFFFFFFFF

%%%%%%%%%%%%%%%   Figure 4       %%%%%%%%%%%%%%%%%   FFFFFFFFFFFFFFFFFFFFFFFFFFFFFFFFFFFFFF

\begin{figure}
\epsfysize=7cm \hspace{2 cm}\epsfbox{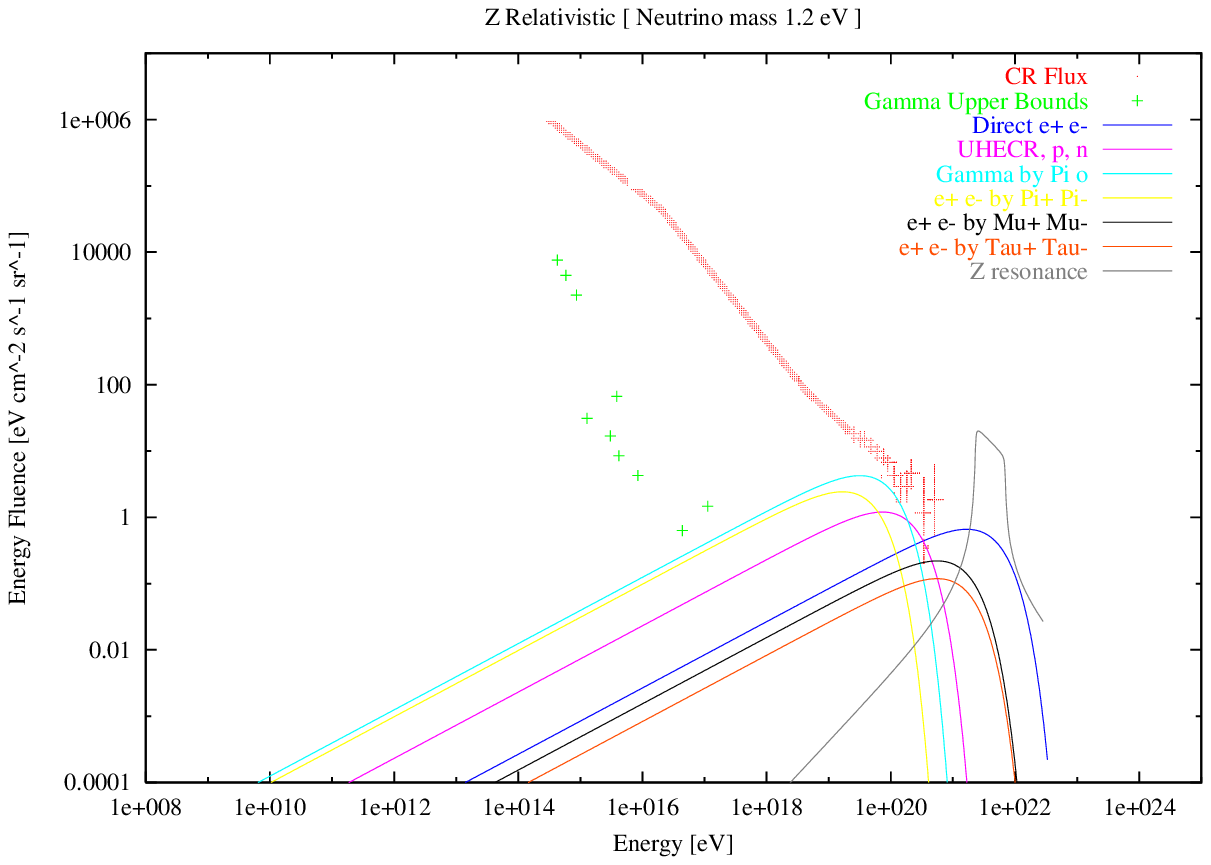} \caption[h]{Energy
Fluence derived by $\nu \bar{\nu} \rightarrow Z$ and its
showering into
  different channels as in previous Figure 2: direct electron pairs UHECR nucleons $n$ $p$, $\gamma$ by $\pi^0$ decay,
  electron pair by $\pi^+ \pi^-$ decay, electron pairs by direct muon and tau decays as labeled in figure.
  In the present case the relic neutrino mass has been assumed to be fine tuned to explain GZK UHECR tail:
  $m_{\nu}=1.2 eV$ with the same UHE incoming neutrino fluence of previous figure. The Z resonance curve shows the averaged $Z$ resonant cross-section peaked
  at $E_{\nu}=3.33\cdot10^{21} eV$.Each channel shower has been normalized in analogy to table 1B.}
\end{figure}

%%%%%%%%%%%%%%%   Figure 4 END   %%%%%%%%%%%%%%%%%   FFFFFFFFFFFFFFFFFFFFFFFFFFFFFFFFFFFFFF
We remind again that a heavier neutrino mass $(\geq 2 eVs)$ imply
the rise of WW-ZZ channels and a pile up of Z resonance
cross-section at lower UHECR spectra. This feature maybe already
responsible for the tiny bump in observed events around
$5\cdot10^{19}eV$. The lighter neutrino mass  possibilities (near
0.1 eV) are comparable with present Super-Kamiokande atmospheric
neutrino mass and are leading to the exciting scenario where more
non degenerated Z-resonances occur (Fargion 2001). These scenario
are summarized in Fig. 5 (for nominal example $m_{\nu_{\tau}}$ =
0.1 eV; $m_{\nu_{\mu}}$ = 0.05 eV). The twin neutrino mass inject
a corresponding twin bump at highest energy. Another limiting case
of interest takes place when the light neutrino masses are
extreme, nearly at atmospheric (SK,K2K) and solar (SNO) neutrino
masses. This case is described in two different versions in Fig.6
(assuming comparable neutrino densities) and Fig.7 (keeping care
of the lightest neutrino density diluitions). The relic neutrino
masses are assumed $m_{\nu_{\tau}}$ = 0.05 eV; $m_{\nu_{\mu}}$ =
0.001 eV). A more complex scenario is also possible when it takes
place both a narrow twin bump (Fig5) $and$ a wider twin bump (Fig
6-7) because of a small neutrino tau-muon mass splitting
overlapping with a wider one due to lightest neutrino electron
mass.

%%%%%%%%%%%%%%%   Figure 4       %%%%%%%%%%%%%%%%%   FFFFFFFFFFFFFFFFFFFFFFFFFFFFFFFFFFFFFF
\begin{figure}
\epsfysize=7cm \hspace{2 cm}\epsfbox{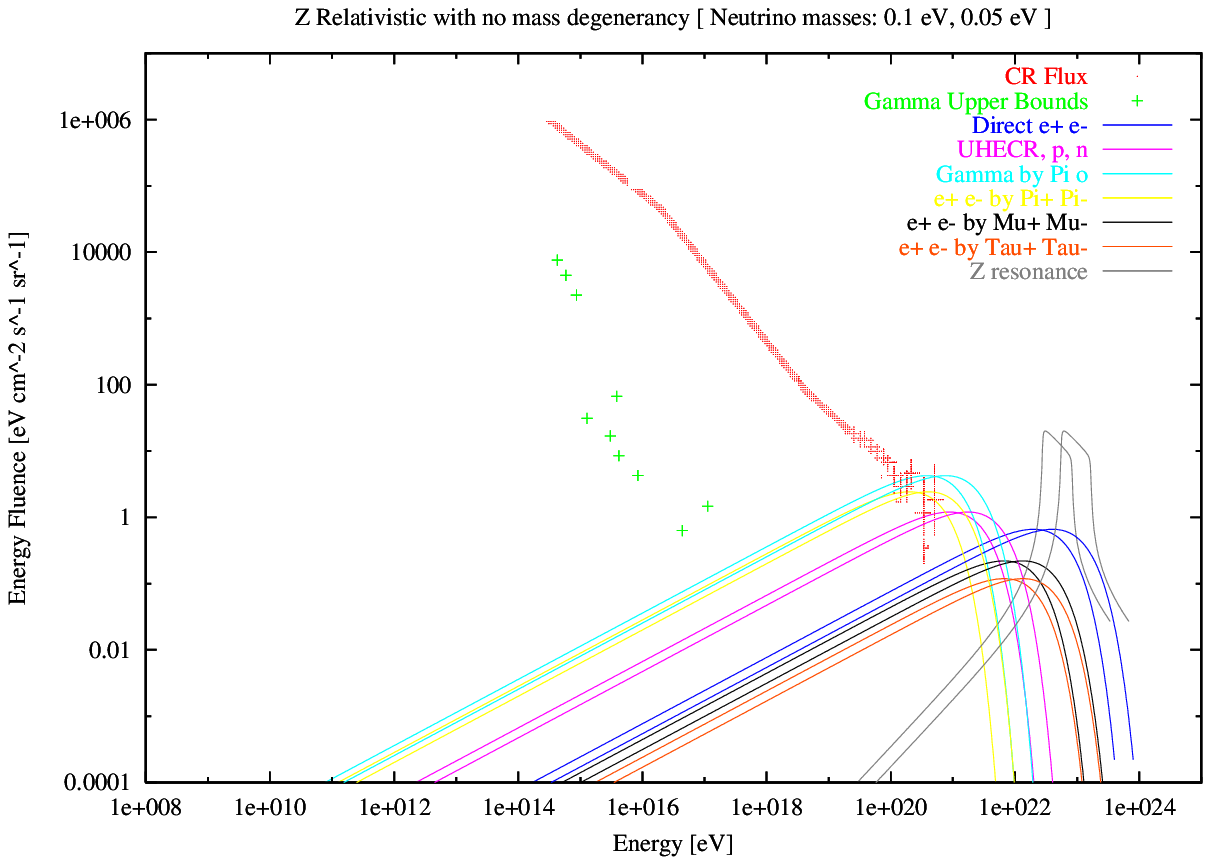} \caption[h]{
Energy Fluence derived by $\nu \bar{\nu} \rightarrow Z$ and its
showering into
  different channels: direct electron pairs UHECR nucleons $n$ $p$, $\gamma$ by $\pi^0$ decay,
  electron pair by $\pi^+ \pi^-$ decay, electron pairs by direct muon and tau decays as labeled in figure.
  In the present case the relic neutrino masses have been assumed with no degenerancy.
  Their values have been fine tuned to explain GZK UHECR tail:
   $m_{\nu_1}=0.1 eV$ and $m_{\nu_2}=0.05 eV$. No relic neutrino
   density difference has been assumed.
   The incoming UHE neutrino fluence has been increased
   by a factor 2 respect previous Fig.3-4. The Z resonance curve shows the averaged $Z$ resonant cross-section peaked
  at $E_{\nu_1}=4\cdot10^{22} eV$ and $E_{\nu_2}=8\cdot10^{22} eV$. Each channel shower has been normalized in analogy to table 1B.}
\end{figure}

%%%%%%%%%%%%%%%   Figure 4 END   %%%%%%%%%%%%%%%%%   FFFFFFFFFFFFFFFFFFFFFFFFFFFFFFFFFFFFFF

%%%%%%%%%%%%%%%   Figure 5       %%%%%%%%%%%%%%%%%   FFFFFFFFFFFFFFFFFFFFFFFFFFFFFFFFFFFFFF

\begin{figure}
\epsfysize=7cm \hspace{2 cm}\epsfbox{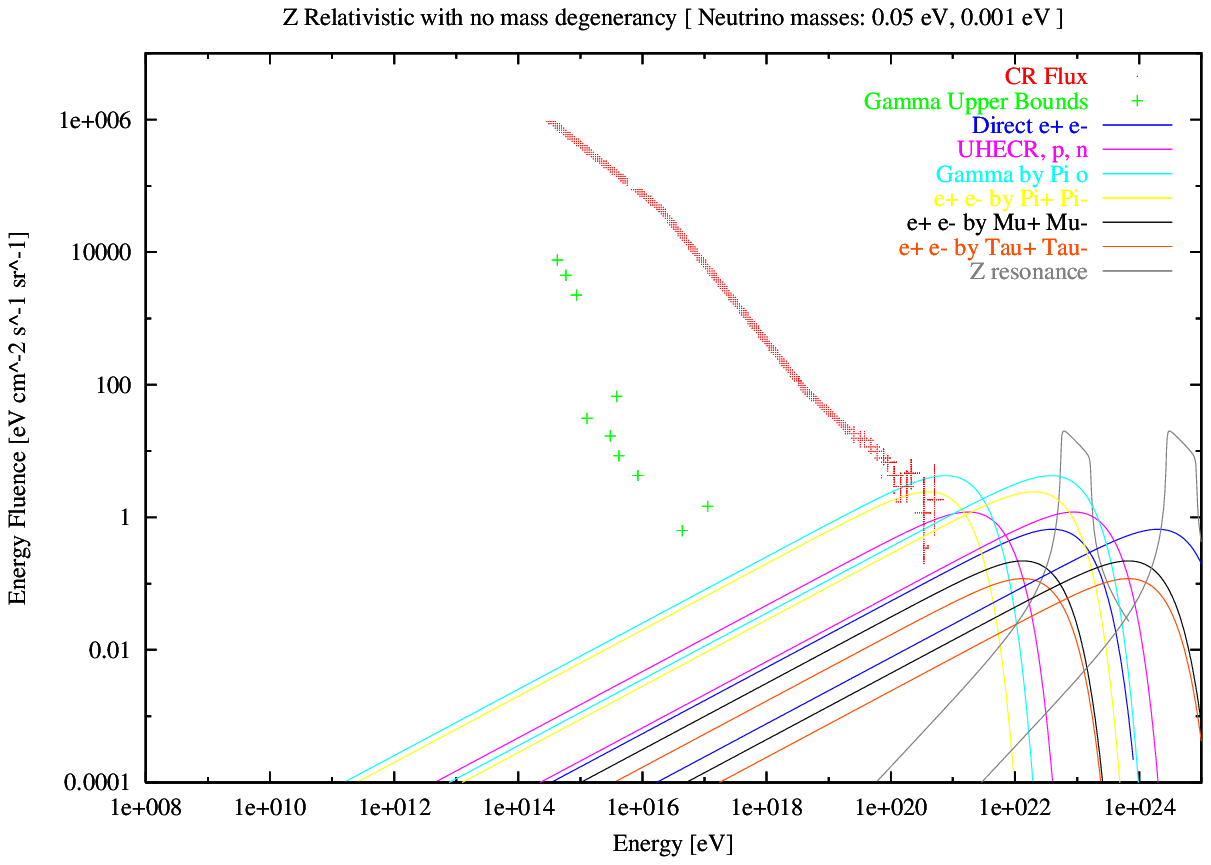}
\caption[h]{Energy Fluence derived by $\nu \bar{\nu} \rightarrow
Z$ and its showering into
  different channels  as above.
  In the present extreme case the relic neutrino masses have been assumed with wide mass differences
  just compatible both with Super-Kamiokande and relic $2 K^{o}$ Temperature.
  The their values have been fine tuned to explain observed GZK- UHECR tail:
   $m_{\nu_1}=0.05eV$ and $m_{\nu_2}=0.001 eV$. No relic neutrino
   density difference between the two masses  has been assumed,
   contrary to bound in eq.3. The incoming UHE neutrino fluence has been increased
   by a factor 2 respect previous Fig.2-3. The "Z resonance" curve
    shows the averaged $Z$ resonant cross-section peaked
  at $E_{\nu_1}=8\cdot10^{22} eV$ and $E_{\nu_2}=4\cdot10^{24} eV$, just
  near Grand Unification energies. Each channel shower has been normalized in analogy to table 1B.}
\end{figure}
%%%%%%%%%%%%%%%   Figure 5 END   %%%%%%%%%%%%%%%%%   FFFFFFFFFFFFFFFFFFFFFFFFFFFFFFFFFFFFFF

%%%%%%%%%%%%%%%   Figure 6       %%%%%%%%%%%%%%%%%   FFFFFFFFFFFFFFFFFFFFFFFFFFFFFFFFFFFFFF

\begin{figure}
\epsfysize=7cm \hspace{2 cm}\epsfbox{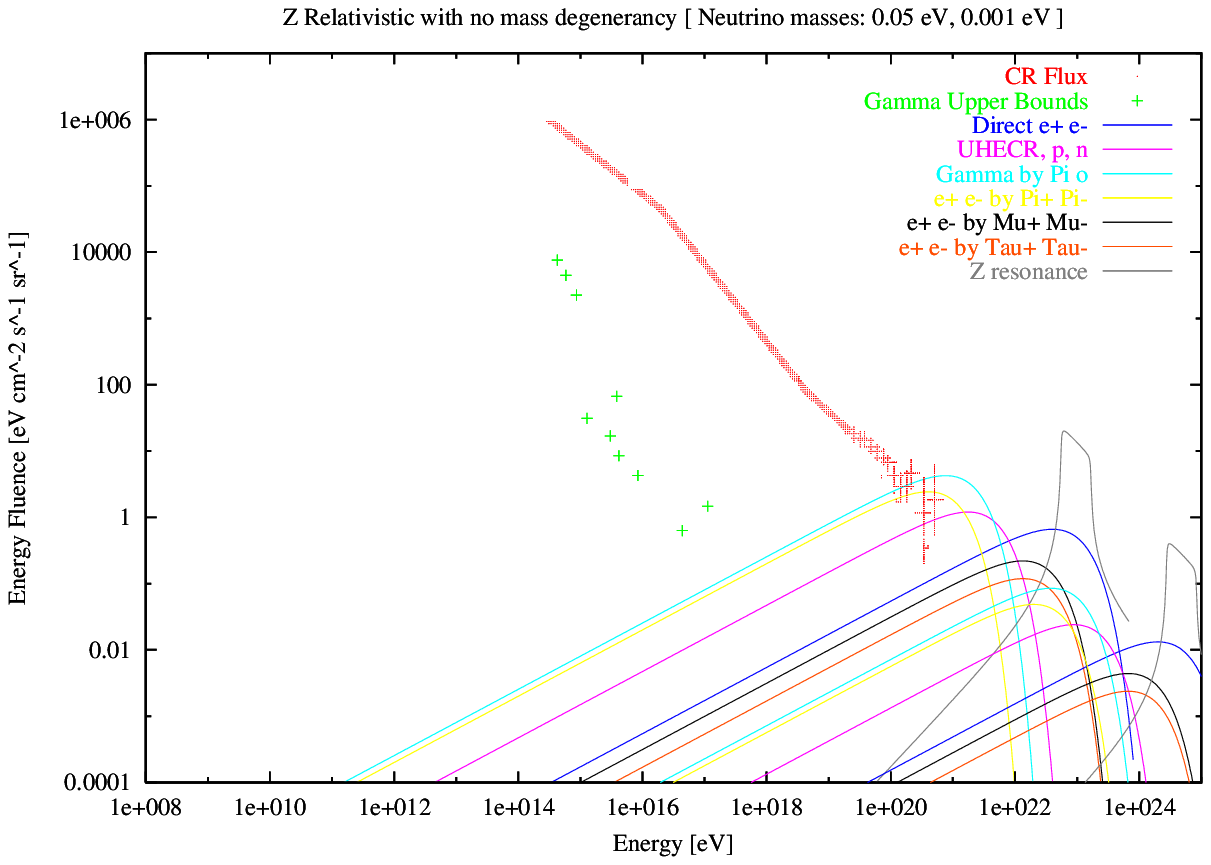}
\caption[h]{Energy Fluence derived by $\nu \bar{\nu} \rightarrow
Z$ and its showering into  different channels  as above.
  In the present extreme case the relic neutrino masses have been assumed with wide mass differences
  just compatible both with Super-Kamiokande and relic $2 K^{o}$ Temperature.
  The their values have been fine tuned to explain observed GZK- UHECR tail:
   $m_{\nu_1}=0.05eV$ and $m_{\nu_2}=0.001 eV$. A neutrino
   density difference between the two masses  has been
   assumed,considering the lightest $m_{\nu_2}=0.001 eV$ neutrino
   at relativistic regime, consistent to bound in eq.3.
   The incoming UHE neutrino fluence has been assumed growing
   linearly (Yoshida et all. 1998)  with energy. Its value is increased
   by a factor 2 and 20  at   $E_{\nu_1}=8\cdot10^{22} eV$ and $E_{\nu_2}=4\cdot10^{24} eV$
   respect the previous ones Fig.2-3. The "Z resonance" curve
    shows its averaged $Z$ resonant "ghost" cross-section peaked
  at $E_{\nu_1}=2\cdot10^{23} eV$ and $E_{\nu_2}=4\cdot10^{24} eV$, just
  near Grand Unification energies. Each channel shower has been normalized in analogy to table 1B.}
\end{figure}

%%%%%%%%%%%%%%%   Figure 6END   %%%%%%%%%%%%%%%%%   FFFFFFFFFFFFFFFFFFFFFFFFFFFFFFFFFFFFFF

\section{UHECRs Clustering and Anisotropy  }

The  neutrino mass play a role in defining its Hot Dark Halo size
and the consequent enhancement of UHECR arrival directions due to
our peculiar position in the HDM halo. Indeed for a heavy $\geq 2
eV$ mass case HDM neutrino halo are mainly galactic and/or local,
reflecting an isotropic or a diffused amplification toward nearby
$M31$ HDM halo. In the lighter case the HDM should include the
Local Cluster up to Virgo. To each size corresponds also a
different role of UHECR arrival time. The larger the HDM size the
longer the UHECR random-walk travel time (in extra-galactic
random magnetic fields) and the wider the arrival rate lag
between doublets or triplets. The smaller is the neutrino halo
the earlier the UHE neutron secondaries by Z shower will play a
role: indeed at $E_{n}= 10^{20}eV$ UHE neutron are flying a Mpc
and their directional arrival (or their late decayed proton
arrival) are more on-line toward the source. This may explain the
high self collimation and auto-correlation of UHECR discovered
very recently (Tinyakov et all. 2001). The UHE neutrons
Z-Showering fits with the harder spectra observed in clustered
events in AGASA (Takeda et all.2001). The same UHECR alignment may
explain the quite short (2-3 years)(Takeda 2001) lapse of time
observed in AGASA doublets. Indeed the most conservative scenario
where UHECR are just primary proton from nearby sources at GZK
distances (tens of Mpcs) are no longer acceptable either because
the absence of such nearby sources and because of the observed
stringent UHECR clustering ($2^o - 2.5^o$) (Takeda et all. 2001)
in arrival direction, as well as because of the short ($\sim3$
years) characteristic time lag between clustered events. Finally
the same growth with energy of UHECR neutron (and anti-neutron)
life-lengths (while being marginal or meaning-less in tens Mpcs
GZK flight distances) may naturally explain, within a $\sim Mpc $
Z Showering Neutrino Halo, the arising harder spectra revealed in
doublets-triplet spectra (Aoki et all. 2001).

\section{The apparent Tinyakov-Glushkov  Paradox }

The same role of UHE neutron secondaries from Z showering in HDM
halo may also solve an emerging puzzle: the  correlations of
arrival directions of UHECRs found recently (Glushkov et all.
2001) in Yakutsk data at energy $E= 8\cdot 10^{18} eV$ toward the
Super Galactic Plane are to be compared with the compelling
evidence of UHECRs events ($E= 3\cdot 10^{19} eV$ above GZK)
clustering toward well defined BL Lacs at cosmic distances
(redshift $z> 0.1-0.2$) (Tinyakov et Tkachev 2001; Tinyakov et al.
2001). Where is the real UHECR sources location? At
Super-galactic disk (50 Mpcs wide, within GZK range) or at cosmic
($\geq 300Mpcs$) edges? It should be noted that even for the Super
Galactic hypothesis (Glushkov et all. 2001) the common proton are
unable to justify the high collimation of the UHECR events. Of
course both results (or just one of them) maybe a statistical
fluctuation. But both studies seem statistically significant
(4.6-5 sigma) and they seem in obvious disagreement. There may be
still open the possibility of $two$ new categories of UHECR
sources both of them located at different distances above GZK
ones (the harder the most distant BL Lac sources). But it seem
quite unnatural  the UHECR propagation by direct nucleons   where
the most distant are the harder. However our Z-Showering scenario
offer different solutions: (1) The Relic Neutrino Masses define
different Hierarchical Dark Halos and privileged arrival direction
correlated to Hot Relic Neutrino Halos. The real sources are at
(isotropic) cosmic edges (Tinyakov et Tkachev 2001; Tinyakov et
al. 2001), but their crossing along a longer anisotropic relic
neutrino cloud enhance the interaction probability in the Super
Galactic Plane. (2) The nearest SG sources are weaker while the
collimated BL Lacs are harder: both sources need a Neutrino Halo
to induce the Z-Showering UHECRs. More data will clarify better
the real scenario.

\section{ The TeV Tails from UHE electrons }

 As it is shown in Table 1A-B and Figures above, the electron
(positron) energies by $\pi^{\pm}$ decays is around $E_e \sim 2
\cdot 10^{19} \, eV$ for an initial $E_Z \sim 10^{22} \, eV $ (
and $E_{\nu} \sim 10^{22} \, eV $). Such electron pairs while not
radiating efficently in extra-galactic magnetic fields will be
interacting with the galactic magnetic field ($B_G \simeq 10^{-6}
\,G $)  leading to direct TeV photons:
\begin{displaymath}
  E_{\gamma}^{sync} \sim \gamma^2 \left( \frac{eB}{2\pi m_e } \right)
   \sim
\end{displaymath}
\begin{equation}\label{4b}
  \sim 27.2 \left( \frac{E_e}{2 \cdot10^{19}
  \,eV} \right)^2 \left( \frac{m_{\nu}}{0.4 \, eV} \right)^{-2} \left( \frac{B}{\mu G} \right)\,TeV.
\end{equation}
The same UHE electrons will radiate less efficiently with extra-
galactic magnetic field ($B_G \simeq 10^{-9} \,G $)  leading also
to direct peak $27.2$ GeV  photons.
   The spectrum of these photons is characterized by a power of law $dN
/dE dT \sim E^{-(\alpha + 1)/2} \sim E^{-1.25}$ where $\alpha$ is
the power law of the electron spectrum, and it is showed in
Figures above. As regards the prompt electrons at higher energy
($E_e \simeq 10^{21}\, eV$), in particular in the t-channels,
their interactions with the extra-galactic field first and
galactic magnetic fields later is source of another kind of
synchrotron emission around tens of PeV energies
$E^{sync}_{\gamma}$:

\begin{equation}\label{2}
 % E^{sync}_{\gamma}
 \sim
  6.8 \cdot 10^{13} \left( \frac{E_e}{10^{21}\,eV} \right)^2
  \left( \frac{m_{\nu}}{0.4 \, eV} \right)^{-2} \left( \frac{B}{nG}
  \right) \, eV
\end{equation}
\begin{equation}\label{2b}
 % E^{sync}_{\gamma}
 \sim
  6.8 \cdot 10^{16} \left( \frac{E_e}{10^{21}\,eV} \right)^2
  \left( \frac{m_{\nu}}{0.4 \, eV} \right)^{-2} \left( \frac{B}{\mu G}
  \right) \, eV
\end{equation}
The corresponding energy loss length instead is (O.E.Kalashev,
V.A.Kuzmin, D.V.Semikoz 2000)
\begin{equation}\label{3}
\left( \frac{1}{E} \frac{dE}{dt} \right)^{-1} = 3.8 \times \left(
\frac{E}{10^{21}} \right)^{-1} \left( \frac{B}{10^{-9} G}
\right)^{-2} \, kpc.
\end{equation}
For the first case the interaction lenght is few Kpcs while in
the second one in few days light flight. Again one has the same
power law characteristic of a synchrotron spectrum with index
$E^{-(\alpha + 1 / 2)} \sim E^{-1.25}$.
 Gammas at $10^{16} \div 10^{17}$ eV scatters onto
low-energy photons from isotropic cosmic background ($\gamma + BBR
\rightarrow e^+ e^-$) converting their energy in electron pair.
 The expression of the pair production cross-section is:
\begin{equation} \sigma (s) = \frac{1}{2} \pi r_0^2 (1 - v^2) [
(3 - v^4) \ln \frac{1 + v}{1 - v} - 2 v (2 - v^2) ]
\end{equation}
where $v = (1 - 4m_e^2 / s)^{1/2}$,  $s = 2 E_{\gamma} \epsilon (1
- \cos \theta)$ is the square energy in the center of mass frame,
$\epsilon$ is the target photon energy, $r_0 $  is the classic
electron radius, with a peak cross section value at
\[ \frac{4}{137}\times \frac{3}{8\pi} \sigma_T \ln 183 = 1.2 \times
10^{-26} \,cm^2 \] Because the corresponding attenuation length
due to the interactions with the microwave background is around
ten kpc, the extension of the halo plays a fundamental role in
order to make this mechanism efficient or not. As is shown in
Fig.3-4, the contribution to tens of PeV gamma signals by Z (or
W) hadronic decay, could be compatible with actual experimental
limits fixed by CASA-MIA detector on such a range of energies.
Considering a halo extension $l_{halo} \gtrsim 100 kpc$, the
secondary electron pair creation becomes efficient, leading to a
suppression of the tens of PeV signal. So electrons at $E_e \sim
3.5 \cdot 10^{16} \,eV$ loose again energy through additional
synchrotron radiation (O.E.Kalashev, V.A.Kuzmin, D.V.Semikozb
2000), with maximum $E_{\gamma}^{sync}$ around
\begin{equation}\label{3b}
  \sim 79 \left( \frac{E_e}{10^{21}
  \,eV} \right)^4 \left( \frac{m_{\nu}}{0.4 \, eV} \right)^{-4}
   \left( \frac{B}{\mu G} \right)^3 \, MeV.
\end{equation}
Anyway this signal is not able to pollute sensibly the MeV-GeV;
the relevant signal pile up at TeVs.

%%%%%%%%%%%%%%%%%%%%%%%%%%%%%%%%%%%%%%%%%%%%%%%%%%%%%%%%%%%%%%%%%%%%%%%%%%%%%%%5
Gamma rays with energies up to 20 TeV have been observed by
terrestrial detector only by nearby sources like Mrk 501 (z =
0.033) or very recently by MrK 421 (z = 0.031). More recent
evidences of tens TeVs from  (three times more) distant blazar
1ES1426+428 (z = 0.129) make even more dramatic the IR-TeV
cut-off. This is puzzling because the extra-galactic TeV spectrum
should be, in principle, significantly suppressed by the
$\gamma$-rays interactions with the extra-galactic Infrared
background, leading to electron pair production and TeVs cut-off.
The recent calibration and determination of the infrared
background by DIRBE and FIRAS on COBE have inferred severe
constrains on TeV propagation. Indeed, as noticed by Kifune
(Kifune 1997), and Protheroe and Meyer (Protheroe et Meyer 2000)
we may face a severe infrared background - TeV gamma ray crisis.
This crisis imply a distance cut-off, incidentally, comparable to
the GZK one. Let us remind also an additional evidence for IR-TeV
cut-off is related to the possible discover of tens of TeV
counterparts of BATSE GRB970417, observed by Milagrito(R. Atkins
et all. 2000), being most GRBs very possibly at cosmic edges, at
distances well above the IR-TeV cut-off ones. In this scenario it
is also important to remind the possibilities that the Fly's Eye
event has been correlated to TeV pile up events in HEGRA (Horns
et al. 1999). The very recent report (private communication 2001)
of the absence of the signal few years  later at HEGRA may be
still consistent with a limited UHE TeV tail activity.
 To solve the IR-TeV cut-off one may alternatively invoke unbelievable extreme hard intrinsic
spectra or exotic explanation as gamma ray superposition of
photons or sacrilegious  Lorentz invariance violation (G.
Amelino-Camelia,et all 1998).
%%%%%%%%%%%%%%%   Figure 7       %%%%%%%%%%%%%%%%%   FFFFFFFFFFFFFFFFFFFFFFFFFFFFFFFFFFFFFF
\begin{figure}
\epsfysize=7cm \hspace{2 cm}\epsfbox{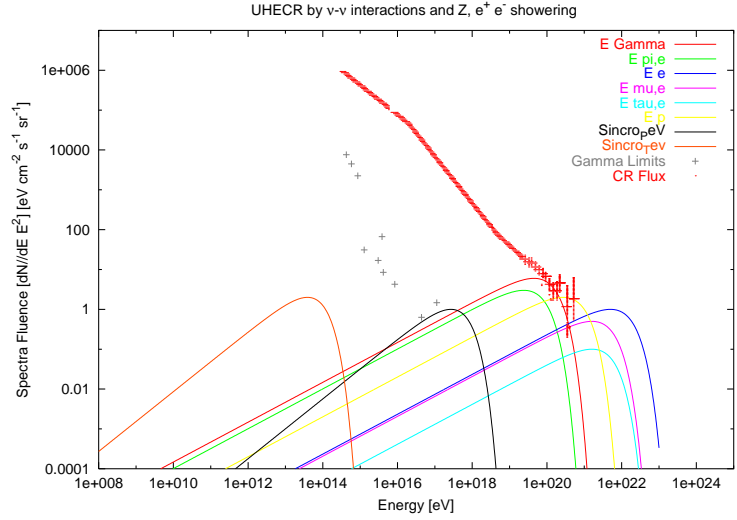} \caption[h]{Energy
fluence by Z showering as in fig.3 for $m_{\nu} = 0.4 eV$ and
$E_{\nu} = 10^{22} eV $and the consequent $e^+ e^-$ synchrotron
radiation by eq.16-18}
\end{figure}

%%%%%%%%%%%%%%%   Figure 7END   %%%%%%%%%%%%%%%%%   FFFFFFFFFFFFFFFFFFFFFFFFFFFFFFFFFFFFFF

%%%%%%%%%%%%%%%   Figure 8       %%%%%%%%%%%%%%%%%   FFFFFFFFFFFFFFFFFFFFFFFFFFFFFFFFFFFFFF
\begin{figure}
\epsfysize=7cm \hspace{2 cm}\epsfbox{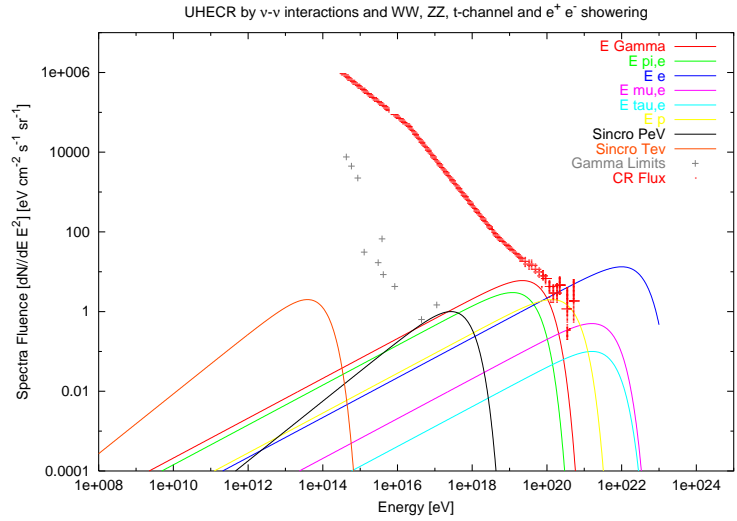} \caption[h]{Energy
fluence by WW, ZZ, t-channel showering as in fig.3, for $m_{\nu}
= 0.4 eV$ and $E_{\nu} = 2 \cdot 10^{22} eV $, and the consequent
$e^+ e^-$ synchrotron radiation  by eq.16-18. The lower energy Z
showering is not included to make spectra more understandable.}
\end{figure}
%%%%%%%%%%%%%%%   Figure 8END   %%%%%%%%%%%%%%%%%   FFFFFFFFFFFFFFFFFFFFFFFFFFFFFFFFFFFFFF

\section{Conclusion}
UHECR above GZK may be naturally born by UHE $\nu$ scattering on
relic ones.  The target cosmic $\nu$ may be light and dense as
the needed ones in HDM model (few eVs). Then their $W^+ W^-,ZZ$
pair productions channel (not just the  Z resonant peak) would
solve the GZK puzzle. At a much lighter, but fine tuned case
$m_{\nu}\sim 0.4 eV$, $m_{\nu}\sim 1.5 eV$ assuming $E_{\nu}\sim
10^{22} eV$, one is able to solve at once the known UHECR data at
GZK edge by the dominant Z peak; in this peculiar scenario one may
foresee  (fig.2-3) a rapid   decrease (an order of magnitude in
energy fluence) above $3\cdot10^{20}eV$ in future data
   and a further recover (due to WW,ZZ channels) at higher energies.
   The characteristic UHECR fluxes will reflect the averaged
   neutrino-neutrino interactions shown in Fig.2-7.
      Their imprint could confirm the neutrino masses value and relic
   density. At a more extreme lighter neutrino mass, occurring for
$m_{\nu}\sim m_{\nu_SK}\sim 0.05 eV$, the minimal
$m_{\nu_{\tau}},m_{\nu_{\mu}}$ small mass differences might be
reflected, in a spectacular way, into UHECR modulation quite above
the GZK edges.  The "twin" lightest masses (Fig.5-6-7) call for
either gravitational $\nu$ clustering above the expected one
 or the presence of relativistic diffused background.
Possible neutrino gray body spectra, out of thermal equilibrium,
at higher energies may also arise from non standard early
Universe. The UHECR acceleration is not yet solved, but their
propagation from far cosmic volumes is finally allowed.
 The role of UHE neutrons in Z-showering, their directional flight leading to clustering in
  self collimated  data is possibly emerging by harder spectra.
 Peculiar secondaries of TeVs tails may be precursor and
 afterglows signal correlated to past or future UHECRs
pointing toward the same far sources. The IR-TeV solution may be
just be a necessary corollary of the Z-Showering GZK solution
(Fargion, Grossi et Lucentini 2001). The  time and space
directional may be a new fundamental test of present  Z-Showering
model. The discover of UHE neutrino at GZK energies might be
testify on ground by UHE $\tau$ air-shower, born by direct
$10^{19}$eV UHE $\nu$ crossing small Earth crust depth, flashing
from the horizontal edges to mountain,balloon and satellite
detectors (Fargion 2000).
 The new generation  UHECR data within next decade,  may also offer the
 probe of lightest elementary particle masses, their
relic densities, their spatial map distribution and energies and
the most ancient and evasive shadows of earliest $\nu$ cosmic
relic backgrounds.

\newpage
\noindent {\bf DISCUSSION} \vskip 0.4cm \noindent {\bf JIM BEALL:}
For photo-pion production in the AGN sources, where do the pions
come from? This is a somewhat rhetorical question.\\

\noindent {\bf D.FARGION:} In the UHE $\nu$-$\nu$ scattering to
the overcome GZK cut-off we consider two places where pions
occurs:
\begin{enumerate}
  \item Near the AGN source: the UHE photon (or nuclei) hitting BBR
(2,7K) photons or local thermal photons lead to:
%\begin {center}
$p+\gamma\rightarrow\triangle^{+}\rightarrow\pi^{+}+n$\\
$p+\gamma\rightarrow\triangle^{+}\rightarrow\pi^{0}+p$\\
or\\
$p+\gamma\rightarrow N\pi+p$\\
$p+\gamma\rightarrow N\pi+n$\\
These UHE pions decays to UHE $\nu$. The UHE $\nu^{+}$ relic
$\nu$ produce Z (UHE parent of UHECR). The Z decay into 2.7
nucleons and $\simeq 30$ pions (neutral + charged).\\ The neutral
pions decay into UHE $\gamma$, the charged ones decay into $\mu$
$\rightarrow e$, and $\nu_{e}$,$\nu_{\mu}$.\\
\end{enumerate}
\noindent {\bf WOLFGANG KUNDT:} What are your objections to a very
near population of CR sources, d$\leq Kpc^{2}$. (I Like to think
of slingshot-acceleration of CR's by neutron-star magnetospheres).\\

\noindent {\bf D.FARGION:} I am sorry because I was first
understanding your "near objects" as Farrar-Bierman-Piran
solution of AGN near (M-87 - like $\leq$ few Mpc) + $B_{G}$. They
should be also UHE neutron, $\bar{n}$
($p+\gamma\rightarrow\pi^{+}+n$ by photopion), source that should
point directly to M-87 (this was not observed). About "galactic"
(Kpc radius) I think they should follow the galactic disk
distribution with strong quadrupole modulation in UHECR
(unobserved). For HALO source  tiny dipole anisotropy might
be hidden but present data seem not correlated with known PSR distribution.\\

\noindent {\bf S.COLAFRANCESCO:} Which is the contribution on
$\Omega_0$ by the relic neutrinos you need ($m_\nu\simeq 0.1\div10
eV$)in the dark halo ($R_{halo} \simeq1\div2 Mpc$) in order to
solve the GZK problem?\\

\noindent {\bf D.FARGION:} The amount of dark matter I consider in
cosmology is linearly dependent on $m_\nu$ mass (for simplicity
here I mean an  unique flavour mass dominant component)
\begin{equation}
%\sqrt{(\Delta m)^2}
{{\Omega_{\nu} \simeq  10^{-2}{\left( \frac{m_{\nu}} {eV} \right)}
 \,} \nonumber}
\end{equation}
The amount of dark matter in galactic or local group halo has a
density which may reach a density contrast:
\begin{equation}
%\sqrt{(\Delta m)^2}
{{\frac{\delta\varrho} {\varrho} \simeq
10^{2}\div10^3{\left(\frac{m_{\nu}} {eV} \right)^\beta}
 \,} \nonumber}
\end{equation}
model dependent  $0\leq\beta\leq2$. The bounds we found up to day
are fixing:

$ m_{\nu}\gtrsim0.1 eV$(extended extragalactic halo);  $2eV  \lesssim m_{\nu} \lesssim 5eV $ (galactic halo).\\
\end{document}